\documentclass{emulateapj}
\usepackage{epsfig}

\def\gsim{\;\rlap{\lower 2.5pt
 \hbox{$\sim$}}\raise 1.5pt\hbox{$>$}\;}
\def\lsim{\;\rlap{\lower 2.5pt
   \hbox{$\sim$}}\raise 1.5pt\hbox{$<$}\;}

\newcommand{\vk}{v_{\mathrm{kick}}}
\newcommand{\ve}{v_{\mathrm{esc}}}
\newcommand{\td}{$t_{\mathrm{dyn}}$}
\newcommand{\tdn}{t_{\mathrm{dyn}}}
\newcommand{\vc}{v_{\mathrm{circ}}}

\begin{document}

\title{ Core Formation in Galactic Nuclei Due to Recoiling Black
Holes}

\author{Michael Boylan-Kolchin\altaffilmark{1}, Chung-Pei
Ma\altaffilmark{2}, and Eliot Quataert\altaffilmark{2}}
\altaffiltext{1}{Department of Physics, University of California,
Berkeley, CA 94720; mrbk@astro.berkeley.edu}
\altaffiltext{2}{Department of Astronomy, University of California,
Berkeley, CA 94720}
\begin{abstract}

Anisotropic gravitational radiation from a coalescing black hole
binary can impart a recoil velocity of up to several hundred km/s to
the remnant black hole.  We examine the effects of recoiling massive
black holes on their host stellar bulges, both for holes that escape
their host and those that return to the galactic center via dynamical
friction.  We show that removal of a black hole via radiation recoil
generally results in a rapidly-formed central core in the stellar
system, with the effect being largest when the hole stays bound to the
bulge and the recoil velocity is comparable to the bulge velocity
dispersion.  Black hole recoil therefore provides a mechanism for
producing cores in some early type galaxies, but it is expected to be
most efficient in faint ellipticals that are known to have steep
density profiles.  We argue that these results may hint at a
significant role for gas in facilitating the coalescence of binary
black holes in faint (power-law) early-type galaxies.

\end{abstract}

\section{Introduction}

When galaxies with central black holes merge, dynamical
friction will drive the holes toward the center of the remnant,
creating a black hole binary.  If the binary separation $a_b$ becomes
small enough that gravitational wave emission is significant, the
binary will rapidly coalesce.  The primary mechanism for reducing
$a_b$ is three-body interactions: stars with orbits passing close to
the binary are scattered to high velocities, allowing the binary's
orbit to decay \citep{bbr80}.  The timescale for removing stars from
low angular momentum orbits (the ``loss cone'') via gravitational
slingshot is much shorter than the (collisional) timescale for
repopulating the loss cone in a spherical galaxy \citep{makino04}.
The latter is many 
Gyr or longer \citep{mm03}, implying that the decay of the binary
would stall before gravitational waves become important.

Recent work has found that a number of processes can increase the
hardening rate: Brownian motion of the binary \citep{quinlan97,
chat03}, triaxiality of the stellar bulge \citep{yu02}, interaction
with a massive accretion disk (Armitage \& Natarajan 2002), and
re-ejection of previously ejected stars \citep{mm03} all may help the
binary coalesce.  Coalescing black hole binaries radiate
gravitational waves that carry away energy and both angular and linear
momentum.  The remnant black hole then receives a recoil velocity
$\vk$ in the range of 100 to 500 km/s \citep{favata04}.  The black
hole can therefore either leave the host galaxy (if $\vk > \ve$) or
oscillate around the center of the galaxy, its orbit decaying due to
dynamical friction. Radiation recoil has a number of astrophysical
implications, from constraining the growth of high-redshift quasars to
the possibility of an intergalactic population of massive black holes
\citep{madau04, merritt04, haiman04}.  In this paper we use numerical
experiments to study the reaction of stellar bulges to coalescing
black holes at their centers (also see \citealt{merritt04}).

\section{Simulations}

We perform N-body simulations to follow the evolution of a stellar
bulge containing a central supermassive black hole that has been given
a recoil velocity immediately following a binary coalescence.  The
(purely gravity) simulations are performed using {\tt GADGET}
\citep{gadget}, a publicly available N-body tree code.  We generate
initial conditions using the equilibrium distribution function for a
spherical stellar system with a central black hole \citep{tremaine94}.
Since dark matter contributes a small fraction of the mass in the
inner tens of parsecs of a typical elliptical galaxy, we ignore its
contribution in our simulations.  We use the \citet{hernquist90} model
for the stellar system; this model is attractive because of its simple
form and the fact that in projection it closely resembles the de
Vaucouleurs $\mathrm{R}^{1/4}$ surface brightness law.  Furthermore,
energy transfer from orbital decay of a black hole binary to
surrounding stars is expected to transform dense cusps associated 
with the growth of a black hole to profiles no steeper than
$\rho \propto r^{-1}$ \citep{mm01}, so an $r^{-1}$ cusp serves as an
upper limit for the expected stellar distribution around a coalescing
black hole binary.

The density profile of the Hernquist model is given by
$$\rho(r)=\frac{M_*}{2 \pi a^3} \frac{a}{r} \frac{1}{(1+r/a)^3}$$ 
with total mass $M_*$ and scale radius $a$, related to the half-mass
radius by $r_{1/2}=(1+\sqrt{2}) a$.  The escape velocity from the
center of the system and the dynamical time and circular velocity at
the scale radius are given by: 
$ \ve = (2GM_*/a)^{1/2}$, $\tdn \equiv
\tdn (r=a) = (3 \pi^2 a^3/ GM_*)^{1/2}$, and $\vc \equiv \vc (r=a)
=(GM_*/4a)^{1/2}$.
We consider only spherically symmetric, isotropic models, in which
case the phase-space distribution function depends on energy alone and
can be calculated from Eddington's formula \citep{bt87}:
$$f(E)=\frac{1}{\sqrt{8} \pi^2} \int_0^E \frac{d^2 \rho}{d \psi^2}
\frac{d \psi} {\sqrt{E-\psi}}.$$ 
Here $\psi$ is the negative of the gravitational potential,
$$\psi = \frac{GM_*}{r+a}+ \frac{GM_{BH}} {r},$$ and $E$ is the
binding energy per unit mass ($E \ge 0$).  We note that without a
central black hole ($M_{BH}=0$), the only accessible energies are $0
\le E \le GM_*/a$ and $f(E)$ is analytic.  Adding a black hole allows
particles to have $E > GM_*/a$ and $f(E)$ must be computed
numerically.  We initialize particle positions from the mass profile,
and then obtain each particle's energy from the calculated distribution
function: $P(E|r) \propto f(E) \sqrt{\psi(r)-E}$.

In the presence of a central point mass, $f(E)$ is not a monotonic
function of energy (see \citealt{tremaine94} for plots of $f(E)$ for
several values of $M_{BH}/M_*$).  Stability to radial perturbations
must therefore be determined numerically.  We have tested that our
simulations (without black hole recoil) indeed maintain equilibrium
for many dynamical times.  Such systems are expected to evolve
on the two-body relaxation timescale \citep{spitzer71}, forming a
$\rho \propto r^{-7/4}$ cusp on small scales \citep{bw76, preto04},
but the relaxation time is of order $10^{10}$ years or longer in a
typical elliptical galaxy, much longer than the timescales relevant to
this paper.

We set the black hole to stellar bulge mass ratio to be
$M_{BH}/M_*=1/300$ and simulate cases with $\vk$ both above and below
$\ve$ for comparison.  We use $N=10^6$ equal-mass particles to
represent the stellar bulge, and a force softening of
$\epsilon=0.0176\,a$ that allows us to resolve down to $\sim 0.2 r_h$,
where $r_h$ is the sphere of influence of the black hole given by
$M(<r_h) \equiv 2 M_{BH}$ (a definition equivalent to the standard
$r_h=GM_{BH}/ \sigma_*^2$ for a singular isothermal sphere).  $r_h$ is
typically a small fraction of the scale radius of the stellar system,
e.g., $r_h=0.089\,a$ for $M_{BH}/M_*=1/300$.  Since gravity is the
only physics in the simulations, our results can be interpreted at
different length and mass scales by rescaling: for $a \rightarrow
\lambda a$ and $M \rightarrow \beta M$, take $v \rightarrow
(\beta/\lambda)^{1/2} \,v$ and $t \rightarrow (\lambda^3 /
\beta)^{1/2} \,t$.  For reference, a model with stellar mass $4 \times
10^8 M_{\odot}$ and a scale radius of $a=42.5$ pc has \td=1.15 Myr,
$\vc=100$ km/s, $\ve=283$ km/s, and a projected central velocity
dispersion at one-eighth the effective radius $R_E$ of $\sigma_p(R_E/8)
\approx 70$ km/s.

\section{Stellar Density Profiles}

\begin{figure}
\begin{center}
\includegraphics[scale=0.5]{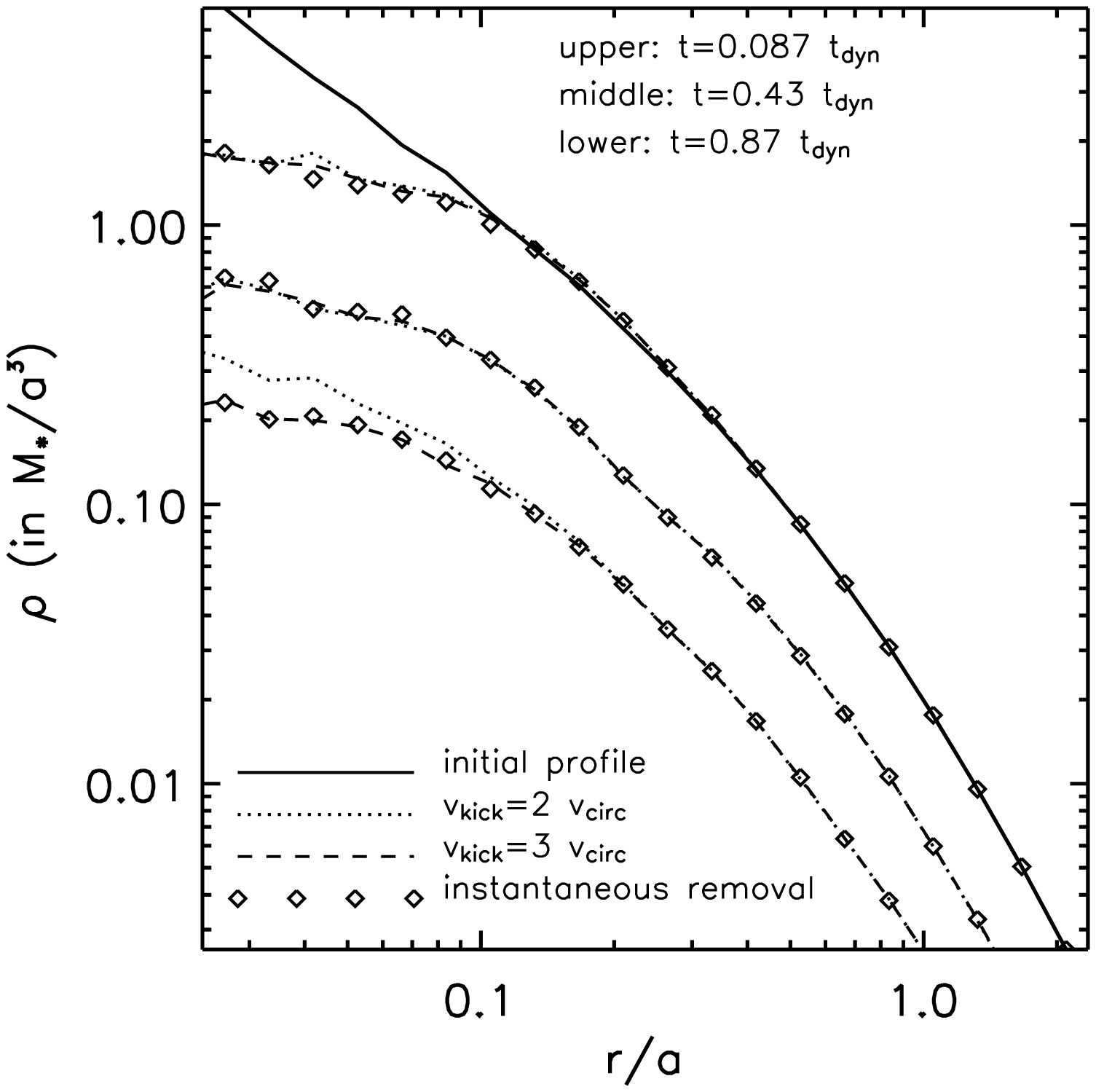}
\vspace{2mm}
\caption{Early evolution of the stellar density profile (in units of
$M_*/a^3$) after the
black hole (with $M_{BH}/M_*=1/300$) recoils (at $t=0$).  Three
simulations are compared: recoil velocities above (dashed) and below
(dotted) the bulge escape velocity $\ve=2.83\,\vc$, and a test model
in which the hole is removed instantly at $t=0$ (diamond).  All three
runs produce a density core inside $\sim r_h$ within $0.087\tdn$ after
the recoil; $\rho$ evolves little afterwards, so for clarity we offset
$\rho$ at 0.43 and 0.87 $\tdn$ by a factor of 3 and 8.  For reference, if
$M_*=4 \times 10^8 M_{\odot}$ and $a=42.5$ pc, \td=1.15 Myr and
$r_h=0.089\,a=3.78$ pc.  }
\label{rho2}
\end{center}
\end{figure}

A consequence of gravitational radiation recoil is evolution of the
density profile $\rho(r)$ of the stellar system.  Fig.~\ref{rho2}
compares $\rho(r)$ at three epochs soon after the recoil of the black
hole.  Within $0.1\,\tdn$ after the black hole leaves the central
region of the bulge, $\rho(r)$ flattens substantially.  A core of
$\sim r_h$ forms and remains for the entire length of the simulation.

Three primary mechanisms could contribute to the rapid flattening of
the inner stellar density profile: (1) departure of stars bound to the
black hole; (2) energy deposition due to dynamical friction on the
black hole; and (3) re-equilibration of the stellar system due to the
departure of the black hole.  We discuss each one in turn.  

In process 1, some stars move with the black hole as it recoils from
the galactic nucleus.  The mass of stars $M_{b}$ bound to the black
hole is a strongly decreasing function of the recoil velocity: for
$\vk /\vc=(0, 0.75, 1, 2, 3)$, our simulations yield $M_{b}/M_{BH}=$
(0.769, 0.394, 0.249, 0.0157, 0.0012).  The removal of stars bound to
the recoiling black hole is therefore too small to account for the
mass deficit of $\sim 2 M_{BH}$ seen in the simulations.

Dynamical heating due to energy transferred to stars from the decay of
the black hole's orbit (process 2) is relatively slow.  The heating
timescale is $\sim |{E / \dot{E}}|$, which is about $5.7\,\tdn$ (in
the $\vk=2\,\vc$ run) given that the change in the black hole energy
from start to apocenter is $\Delta E / E_{t=0}\approx -0.076$ and that
the time required for the black hole to reach apocenter is $\approx
0.43\,\tdn$.  In comparison, Fig.~\ref{rho2} shows core development
over a much shorter timescale ($< 0.1\,\tdn$), indicating that
dynamical friction heating is not the dominant process in the
\emph{initial} core formation.  Its longer term effect will be
discussed below.

\begin{figure}
\begin{center}
\includegraphics[scale=0.5]{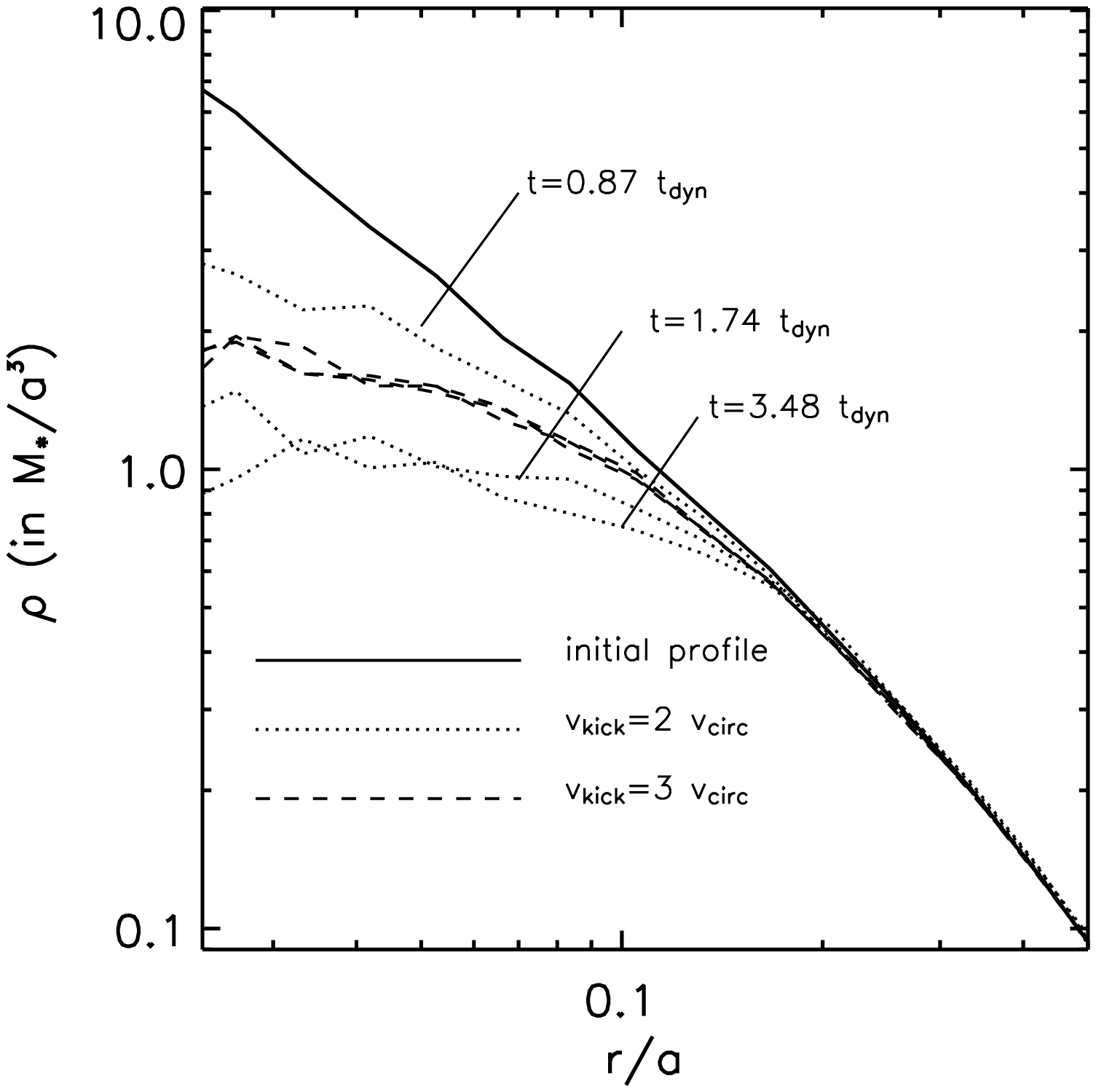}
\vspace{2mm}
\caption{Evolution of the stellar density profile at 0.87, 1.74, and
  $3.48\,\tdn$ for the same runs as in Fig.~1.  For $\vk=3\,\vc
  >\,\ve$, the density profile remains constant once the hole leaves
  the galactic nucleus, while for $\vk=2\,\vc < \ve$ there is both
  early evolution as the stellar bulge dynamically adjusts to the
  absence of the hole and later evolution due to dynamical friction
  heating as the black hole returns to the center of the stellar
  system.}
\label{rho1}
\end{center}
\end{figure}

To study the effects due to process 3, we first note that the
timescale for the stellar system to re-adjust after the black hole
ejection is the crossing time of the core, which is $\approx
0.075\,\tdn$ for a core of $\approx 0.14 a$ shown in Fig.~\ref{rho2}.
This timescale is comparable to the core formation time of $\la
0.087\,\tdn$; the stellar system therefore has sufficient time to
respond to the black hole kick by producing the observed flattening of
$\rho(r)$.  We test the hypothesis that process 3 dominates the early
evolution of the stellar system by performing a simulation in which we
remove the black hole instantaneously from an initial equilibrium
configuration and then allow the system to evolve to its new
equilibrium (also see \citealt{merritt04}).  This test eliminates
evolution due to gravitational interactions between the black hole and
the stars and mimics the effect of $\vk \gg \ve$.  Fig.~\ref{rho2}
compares the stellar $\rho(r)$ for this model with the $\vk=2\,\vc$ and
$3\,\vc$ simulations.  This figure shows that at early times, all
three simulations behave quite similarly; in fact, $\rho(r)$ is
virtually indistinguishable at both 0.087 and 0.43 \td.  This supports
the hypothesis that dynamical adjustment to the absence of the black
hole is responsible for the initial core formation.

The longer-term evolution of the stellar density distribution is shown
in Fig.~\ref{rho1}.  For the run where the black hole escapes
($\vk=3\, \vc$), $\rho(r)$ is essentially unchanged from $0.43\,\tdn$ up
to the end of our simulation at $3.48\,\tdn$, at which point the black
hole is more than $13a$ from the center of the stellar system.  The
evolution is qualitatively different, however, if the black hole
returns.  For $\vk=2\,\vc$, the core density at 0.87 \td\ is higher
than that with $\vk > \ve$.  This effect is due to the presence of the
black hole (and the stars bound to it) on its return path through the
center of the system.  Subsequently, however, the core becomes less
dense relative to the $\vk=3\,\vc$ run, at both 1.74 and 3.48 \td.
From 3.48 \td\ onwards, $\rho(r)$ of the $\vk=2\,\vc$ run is
essentially unchanged and the black hole is nearly stationary at the
center of the stellar potential.  In both simulations, the velocity
dispersion tensor remains isotropic and the bulge remains mostly
spherical. 

\begin{figure}
\begin{center}
\includegraphics[scale=0.5]{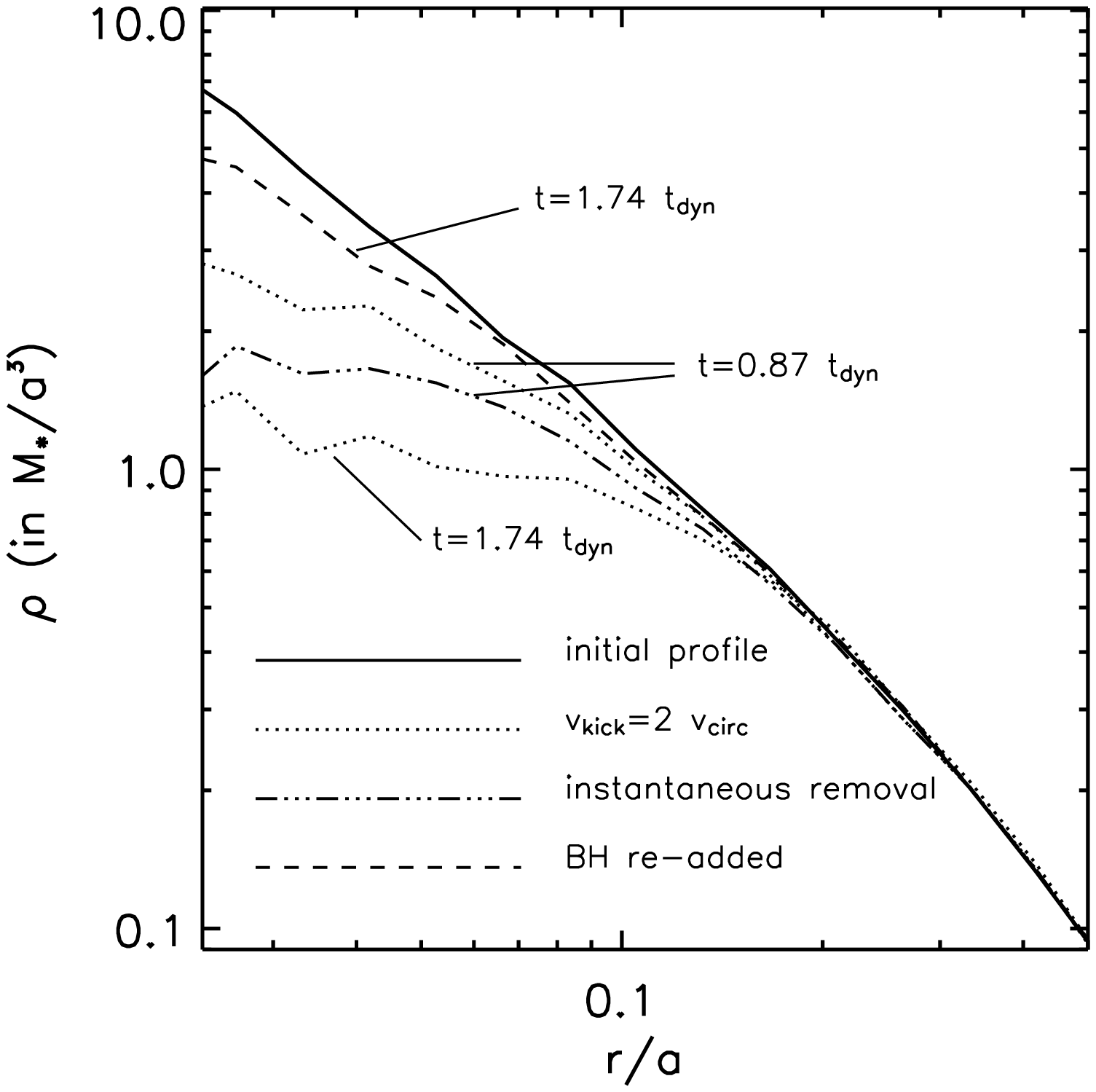}
\vspace{2mm}
\caption{Comparison of the $\vk=2\,\vc < \ve$ simulation (dotted
curves) with a calculation in which the hole is instantaneously
removed at $t=0$ and re-introduced at $0.87\,\tdn$.  The results highlight
the role of dynamical friction: the static re-introduction of a black
hole to the core causes the re-building of a central cusp by
$t=1.74\,\tdn$, while the true simulation shows core-formation by
dynamical friction heating as the hole returns to the galactic
center.}
\label{plot6}
\end{center}
\end{figure}

To explore the origin of the additional flattening in the $\vk=2\,\vc$
calculation, we perform a control simulation in which we re-introduce
the black hole into the ``instantaneous removal'' simulation at 0.87
\td.  In this case, dynamical friction is irrelevant (as in the
instantaneous removal simulation) but the stellar system re-adjusts
because of the black hole.  Fig.~\ref{plot6} shows that the
equilibrium profile attained after re-addition of the black hole is
significantly steeper than that found in the $\vk=2\,\vc$ simulation
at $t=1.74\,\tdn$. This comparison test shows that the additional
flattening seen in the $\vk=2\,\vc$ run is due to the heating of the
stellar system by the black hole as dynamical friction returns it to
the center of the galaxy.  Even though this input of energy is less
than 1/1000 of the total energy of the initial stellar system (for
$\vk=2\,\vc$), it is enough to ensure that the new equilibrium differs
substantially from the original at small radii.

The results of this section show that for recoil velocities $\vk\gsim
\vc$, a steep density cusp is difficult to maintain in a stellar
system containing a coalescing black hole binary (simulations with
$\vk \lsim \vc$ show little change in the density profile of the
bulge).  We now turn to the implications of these results.

\section{Discussion and Implications}

In a hierarchical cosmology, mergers of galaxies
with central black holes will lead to the formation of black hole
binaries.  This process occurs on timescales significantly shorter
than the Hubble time only in the case of major mergers, where the mass
ratio of the two galaxies is within a factor of about 3 \citep{vol03}.
If the timescale for coalescence of the binary black holes is shorter
than the typical time between galaxy mergers, it is likely that
gravitational radiation recoil will have interesting observational
consequences, most notably for early type galaxies.

Observations of early-type galaxies show that their density profiles 
fall into two distinct 
classes \citep{ferrarese94, lauer95}.  Power law galaxies, which tend
to be faint, have surface brightness profiles that are steep and do
not seem to exhibit a significant break at small radii.  Core galaxies
are generally brighter and have a discernible break in their surface
brightness profiles.  Faber et al. (1997) discuss several
possibilities for the origin of core galaxies.  In particular, they
and others have argued that the formation of a black hole binary can
produce a core in a stellar system, both by gravitational slingshot of
stars during the decay of the binary and by Brownian motion of the
binary \citep{ebi91, makino96, quinlan97, faber97, mm01, ravi02,
milo02}.  This process of central density reduction due to
loss cone depletion should be ubiquitous in galaxies with binary black
holes, reducing their central cusps to $\rho \propto r^{-1}$ or
shallower. 

We have demonstrated in this paper that in addition to these
mechanisms for core formation, gravitational recoil in the predicted
velocity range of 100 to 500 km/s generically results in the
formation a core in the density profile of a 
stellar system, which produces a core in the surface brightness
analogous to that seen in bright ellipticals.  (We note that the
late-time {\it volume} density profiles in
Figs.~\ref{rho2}-\ref{plot6} also yield {\it surface} densities that
are nearly flat inside $\approx 1-2\ 
r_h$ but do not affect the projected aperture velocity dispersions
noticeably.)  
For recoil velocities 
$\vk > \ve = 2.83 \vc$ (for the Hernquist model), which are rare for
Milky Way size bulges or larger but plausible for smaller galaxies, we
expect the black hole recoil to produce stellar cores that are largely
independent of $\vk$ since stellar re-equilibration is the dominant
process.  For $\vk < \ve = 2.83 \vc$, however, recoils of a given
$\vk$ have a larger effect on the inner stellar profile in smaller
galaxies because $\vk/\vc$ is larger.  In addition, black holes that
do not escape their host galaxy tend to create larger cores via
dynamical friction than black holes that do escape.  In the context of
these results, it is thus the existence of power law galaxies that
must be explained.  Indeed, one might naively have predicted that
smaller, fainter galaxies would have proportionally larger cores (in
units of $r_h$) because of gravitational recoil, contrary to what is
observed.

One resolution of this difficulty is that black holes are rare in
low-mass galaxies.  The dynamical constraints on black holes in such
galaxies are indeed rather poor, though there are suggestive hints of
substantial population of $\sim 10^5-10^6 M_\odot$ black holes in
dwarf Seyfert galaxies (e.g., \citealt{filippenko03, barth04,
greene04}).  There are also plausible mechanisms that could rebuild a
cusp.  A natural explanation for reforming a power-law surface
brightness profile is star formation accompanying a dissipative merger
of gas-rich galaxies.  This scenario requires  the black hole binary
to coalesce -- and the recoiling hole to return to the galactic
nucleus -- \emph{before} the starburst is completed. 
If stellar dynamical processes alone are responsible for hardening the
binary, the timescale is typically $> 10^8$ years even in the
case of highly triaxial galaxies \citep{yu02}.  This is
comparable to, or somewhat longer than, the expected duration of a
merger-triggered starburst (e.g., \citealt{mihos96}).  An intriguing
possibility is that the presence of substantial amounts of gas can
greatly reduce the time for binary black holes to coalesce.  If this
is the case, as argued by, e.g., Armitage \& Natarajan (2002) and
\citet{escala04}, then it is plausible that a black hole binary can
coalesce before the end of a starburst and a stellar cusp can be
re-formed, thus helping explain the existence of power-law early-type
galaxies that are built hierarchically.  We note that, given this
scenario, in future work it would be interesting to consider the
dynamics of recoiling holes in the presence of a surrounding accretion
disk.

We thank Julie Comerford and Piero Madau for useful discussions and
Volker Springel for making {\tt GADGET} publicly available.  This
research used resources of the National Energy Research Scientific
Computing Center, which is supported by DOE under Contract
No. DE-AC03-76SF00098.  C.-P. M. is supported in part by NSF grant AST
0407351 and NASA grant NAG5-12173.  EQ is supported in part by NSF
grant AST 0206006, NASA grant NAG5-12043, an Alfred P. Sloan
Fellowship, and the David and Lucile Packard Foundation.

\end{document}